
\magnification=\magstep1
\looseness=2
\tolerance 1000
\def\ref{\par\noindent\hangindent 12pt}

\def\msn{\par\nobreak\noindent}
\def\bsn{\bigbreak\goodbreak\noindent}
\def\noi{\noindent}
\def\etal{{\it et al.}\ }
\def\refset{\parindent=0pt\hangafter=1\hangindent=1em}

\hsize=6.25 truein
\vsize=9.0 truein
\baselineskip=25pt plus1pt minus1pt
\parskip=2pt plus1pt minus1pt
\vfill
\centerline{PROBING THE LARGE-SCALE VELOCITY FIELD}
\centerline{WITH CLUSTERS OF GALAXIES}
\vskip 2cm
 \centerline{Neta A. Bahcall,$^1$ Renyue Cen,$^1$ and
Mirt Gramann$^{1,2}$}
\bsn
\centerline{$^1$ Princeton University Observatory, Princeton, NJ 08544}
\centerline{$^2$ Tartu Astrophysical Observatory, T\~oravere, Estonia}
\vskip 2cm
\centerline{\hskip 2cm Received:\hrulefill; Accepted:\hrulefill \hskip 2cm}
\vskip 6cm
\centerline{Submitted to the {\it Ap. J. Letters}}
\vskip 1cm
\centerline{April 27, 1994}
\par\vfill\eject

\centerline {ABSTRACT}
\msn
What is the role of clusters of galaxies in probing the large-scale
velocity field of the universe? We investigate the distribution
of peculiar velocities of clusters of galaxies in the popular
low-density ($\Omega=0.3$) flat Cold-Dark-Matter (CDM) cosmological
model, which
best fits many large-scale structure observations. An $\Omega=1$
CDM model is also studied for comparison. We find that clusters
of galaxies are efficient tracers of the large-scale
velocity field. The clusters exhibit a Maxwellian
distribution of peculiar velocities, as expected from
Gaussian initial density fluctuations. The cluster 3-D
velocity distribution for the $\Omega=0.3$ model
peaks at $v \sim 400$ km s$^{-1}$, and extends to high
velocities of $v \sim 1200$ km s$^{-1}$.
The rms peculiar velocity of the clusters is $440$ km s$^{-1}$.
Approximately 10\% of all model clusters move with high peculiar
velocities of $v \ge 700$ km s$^{-1}$. The observed velocity
distribution of clusters of galaxies is compared with the
predictions from cosmological models. The observed data
exhibit a larger
velocity tail than seen in the model simulations; however, due
to the large observational uncertainties, the data are
consistent at a $\sim 3\sigma$ level with the model predictions,
and with a Gaussian initial density field.  The large peculiar
velocities reported for some clusters of galaxies
($v \geq 3000$ km s$^{-1}$) are likely to be overestimated,
if the current model is viable.

{\it Subject headings: cosmology: theory -- galaxies: clustering}
\par\vfill\eject
\centerline{1. INTRODUCTION}
\msn
Clusters of galaxies are an efficient tracer of the large-scale
structure of the universe. The strong correlation
function of cluster of galaxies (Bahcall \& Soneira 1983; Klypin
\& Kopylov 1983; Postman \etal 1992; Peacock \& West 1992) and
the superclustering properties of clusters (Bahcall \& Soneira 1984;
Postman \etal 1992) provided some of the first evidence
for the existence of organized structure on large scales.

Are clusters of galaxies also useful tracers of the large-scale
peculiar velocity field in the universe? The peculiar velocity
field results most likely from
the gravitational acceleration that develops from initial density
fluctuations in the early universe. The velocity field can
place strong constraints on cosmological models, including the
mass-density of the universe (Bertschinger \& Dekel 1989; Dekel 1994).
Our galaxy moves with $\sim 600$ km s$^{-1}$ relative to the
cosmic frame defined by the Cosmic Microwave Background (CMB),
as implied by the dipole observations of the CMB
(Smoot \etal 1977). Peculiar motions of other
galaxies have been detected by measuring their
velocity deviations from the Hubble flow
by using relative distance indicators that predict the Hubble
velocities of the galaxies (Rubin \etal 1976; Dressler \etal 1987;
Burstein \etal 1987; Faber \etal 1989). The observational
determination of cluster motions can be more accurate than the
determination of peculiar velocities of individual galaxies
since the distance to each cluster is determined from
a large number of member galaxies. Moreover, rich clusters
of galaxies, unlike galaxies themselves, trace the large-scale
linear density field of the universe, thus directly reflecting
the initial conditions.

The possible existence of large peculiar velocities of clusters
in some superclusters, with $v_r \sim 10^3$ km s$^{-1}$, has been
suggested (Bahcall \etal 1986). Using relative
distance indicators, clusters with similarly high peculiar velocities,
$v_r \sim 10^3$ km s$^{-1}$, have been observed (Faber \etal 1989;
Mould \etal 1991, 1993). A large bulk-flow of
relatively nearby rich clusters of galaxies, with
$v_r \sim 700$ km s$^{-1}$, has recently been reported
(Lauer \& Postman 1994).

In this Letter, we investigate the expected motions of clusters
of galaxies in some popular cosmological models.
We first study the velocity distribution of clusters
of galaxies in a low-density CDM model, which best
fits most of the large-scale structure observations, including
the galaxy and the cluster correlation functions, the galaxy
power-spectrum, the mass-function of clusters of galaxies,
the small-scale pairwise velocities of galaxies, as well as
galaxy formation and thermodynamic properties
(Gramann 1988; Maddox \etal 1990; Bahcall \& Cen 1992;
Cen \etal 1993; Kofman \etal 1993). We find that clusters of
galaxies move with considerable speeds; they provide an important
tracer of the large-scale velocities. We compare the
model expectations with observations of group and cluster velocities.
We find that the data exhibit a tail of larger
velocities than predicted by the model; however, within the
large observational uncertainties, the available data are
consistent with the model predictions. An $\Omega=1$ CDM model
yields similar results but with somewhat
larger velocities.
\bsn
\centerline{2. PECULIAR VELOCITIES OF CLUSTERS OF GALAXIES}
\msn
A large-scale Particle-Mesh code with a box size of $800$h$^{-1}$ Mpc
is used to simulate the evolution of the matter.
A large simulation box is needed in order to insure that:
1)~contributions to velocities from waves
larger than the box size are small; and 2)~uncertainties due to
fluctuations in the small number of large waves in the box
are minimized. We find that these conditions are satisfied with
the $800$h$^{-1}$Mpc box size (see below).

The simulation box contains $500^3$ cells and $250^3 = 10^{7.2}$
matter particles. The spatial resolution is $1.6$h$^{-1}$ Mpc.
[A higher resolution ($0.8$h$^{-1}$ Mpc) smaller box
($400$h$^{-1}$ Mpc) is also studied for comparison.]
Details of the simulations are discussed by Cen (1992). A
low-density CDM model with a mass density $\Omega=0.3$,
a cosmological constant $\Omega_{\Lambda}=1-\Omega=0.7$,
a Hubble constant $h=H_{o}/100=0.67$, and a normalization of the
mass fluctuations on $8h^{-1}$ Mpc scale $\sigma_8=2/3$
(as determined by the CMB anisotropy measurements of COBE;
Smoot \etal 1992) is used. For comparison we also investigate
the standard $\Omega=1$ CDM model, with $h=0.5$ and
$\sigma_8=1.05$ (COBE normalization).

Clusters are selected in the simulation using an adaptive linkage
algorithm (Bahcall \& Cen 1992). The threshold for the cluster mass,
within $r=1.5$h$^{-1}$ Mpc of the cluster center, is selected to
correspond to a number density of clusters comparable to the
observed density of rich ($R \ge 1$) clusters, $n_{cl} \sim
6 \times 10^{-6}$ h$^3$ Mpc$^{-3}$ (Bahcall \& Cen 1993). We
also use, for comparison, a lower threshold that corresponds to
the observed number density of small groups of galaxies,
$n_{gr} \sim 10^{-4}$ h$^{3}$ Mpc$^{-3}$ (Ramella, Geller
\& Huchra 1989; Bahcall \& Cen 1993).

A total of $\sim 3000$ rich clusters of galaxies and
$\sim 5 \times 10^4$
groups are obtained in the $800$h$^{-1}$ Mpc simulation box.
The peculiar velocity of each of these clusters
(or groups) relative to the co-moving cosmic frame is
determined from the simulation; these velocities are
used in the analysis described below.

The peculiar velocity distribution of clusters of galaxies,
$P(v)$, is presented in Figure~1a. It represents
the probability distribution, or the normalized number density, of
clusters with peculiar velocities in the range $v \pm dv$, per
unit $dv$, as a function of $v$. The velocity $v$ refers to the
three-dimensional peculiar velocity of the cluster relative to
the cosmic background frame. The solid line represents the velocity
distribution of rich clusters of galaxies (with
$n_{cl}=6 \times 10^{-6}$ h$^{3}$ Mpc$^{-3}$); the dashed line
corresponds to the distribution of the lower-threshold groups of
galaxies (with $n_{gr}= 10^{-4}$ h$^{3}$ Mpc$^{-3}$).
The similarity of the two distributions implies that the cluster
velocity distribution is insensitive to the threshold
(i.e. richness) of the selected clusters.

Figure~1 shows that clusters of galaxies move with significant
speeds. The cluster velocities peak at $v \sim 400$ km s$^{-1}$
(comparable to the velocity of our own Galaxy); their rms peculiar
velocity is 440 km s$^{-1}$. The velocity distribution
exhibits a moderate fall-off at high velocities
($v \sim 500-1000$ km s$^{-1}$), and a
faster drop  at small velocities ($v \le 300$ km s$^{-1}$).
Approximately 35\% of all clusters exhibit
peculiar velocities in the range $v\approx 500 \pm 100$ km s$^{-1}$.
About 10\% of all clusters have peculiar velocities in the range
$v \approx 1000 \pm 300$ km s$^{-1}$; the velocity distribution
tail reaches $\sim 1200$ km s$^{-1}$.

The integrated velocity distribution of clusters of galaxies, $P(>v)$
(representing the probability distribution of clusters with peculiar
velocities larger than $v$), is presented in Figure~1b.
The integrated function emphasizes
the high-velocity tail of the distribution. It reveals that
a significant fraction of all model clusters ($\sim 10$\%) exhibit
high peculiar velocities of $v \ge 700$ km s$^{-1}$;
the tail of the cluster velocity distribution, to
$\sim 1200$ km s$^{-1}$, is clearly seen.

The shape of the cluster velocity distribution is well matched by
a Maxwellian distribution $P(v) \sim v^2 \exp (-v^2/2\sigma^2)$
(dotted line in Figures 1a and 1b). The velocity dispersion,
$\sigma$, is determined from the model simulation such
that the fit and the
model have the same rms velocity. A Maxwellian distribution
is expected from a Gaussian initial density fluctuation field. The
above results suggest that clusters of galaxies provide an important
tracer of the large-scale velocity field and can help to test
the type of the initial density field (Gaussian or non-Gaussian).

How does the velocity distribution of clusters of galaxies depend
on the specifics of the model? We studied an $\Omega=1$ CDM model,
with $h=0.5$ and $\sigma_8=1.05$ (as required for a COBE normalization).
This model, however, unlike the low-density CDM model,
is inconsistent with several large-scale structure observations
such as the number-density and correlation function of clusters
of galaxies and the small-scale pairwise velocities of
galaxies (Maddox \etal 1990; Bahcall \& Cen 1992;
Cen \& Ostriker 1992; Ostriker 1993).
We find that the velocity distribution
of clusters in the $\Omega=1$ CDM model is similar to that of the
$\Omega=0.3$ model but shifted to higher velocities: the peak
velocity is $v \sim 600$ km s$^{-1}$, the velocity tail extends
to $\sim 2000$ km s$^{-1}$, and the rms peculiar velocity
is $762$ km s$^{-1}$ (3D). Both model cluster velocity
distributions are well approximated by a Maxwellian.

The simulation results are consistent with expectations from
linear theory (Gramann \etal 1994). Using linear theory we
find that contributions from waves larger than
$800$h$^{-1}$ Mpc are small; the rms velocities integrated
to $\lambda=800$h$^{-1}$ Mpc and to infinity are $416$ and
$426$ km s$^{-1}$, respectively, for the $\Omega=0.3$ CDM model.
The sensitivity of the results to the resolution of the
simulation is tested by comparing the simulations with a
$0.8$h$^{-1}$ Mpc nominal resolution and with a
$1.6$h$^{-1}$ Mpc
nominal resolution (both in a $400$h$^{-1}$ Mpc box). The
resulting velocity distributions of groups and clusters are
consistent with each other (to within $\leq 60$ km s$^{-1}$)
for both resolutions.
\bsn
\centerline{3. COMPARISON WITH OBSERVATIONS}
\msn
How does the predicted cluster velocity distribution
compare with observations?

Observational determination of peculiar velocities of galaxies, groups,
and clusters is difficult, since the true distances of the objects
and hence their Hubble velocities are uncertain. However, data are
available for the peculiar velocities of some samples of groups
and clusters of galaxies. We use the group and cluster peculiar
velocities observed by the Tully-Fisher (TF) method for distance
indicators (Aaronson \etal 1986; Mould \etal 1991, 1993;
Mathewson \etal 1992), and by the $D_n - \sigma$
method (Faber \etal 1989). Groups with observational
velocity uncertainties
$\ge 900$ km s$^{-1}$ are excluded from the analysis.
A total of 48 group and cluster peculiar velocities are available
from the TF method, and 91 from $D_n - \sigma$. The total
sample of peculiar velocities includes
123 non-overlapping groups and clusters. These data are used to
determine the observed velocity distribution of groups of galaxies.
As seen in the simulations, the velocity distribution is
insensitive to the group threshold: small groups and rich
clusters yield similar results.

The observed differential and integrated group velocity distributions
are presented in Figure 2a-b. Here we use the one-dimensional
velocities, as observed ($v_{1D}$); they are compared with the 1-D
velocities in the models (as opposed to the 3-D velocities used in the
previous discussion, since the 3-D velocities are not directly
observed). We present in different symbols the
data obtained from the TF, $D_n - \sigma$, and total (combined)
samples. The results from the different
sub-samples are approximately consistent with each other; the total
sample is thus used for comparison with model expectations.

The number of rich clusters in the observed sample is small
($\sim 18$ $R \ge 0$ clusters). Within the large statistical
uncertainties of such a small sample, the observed rich cluster velocity
distribution is consistent with that of the groups, and thus consistent
with the model comparisons discussed below.

The observed velocity distribution is superimposed on the 1-D velocity
distribution expected for groups (dotted line; Fig. 2). The shape
of the functions
differs from those of Figures 1 since the 1-D velocity distribution
is plotted instead of the 3-D. The 3-D distribution is proportional
to $v^2$ at small velocities; instead, the 1-D velocities
exhibit a Gaussian distribution, as expected for a 3-D Maxwellian.
In comparing model expectations with observations, the model
velocity distribution (dotted line) is convolved with the observational
uncertainties; each model cluster is given an uncertainty drawn at
random from the {\it actual} distribution of observed uncertainties
of the total sample. The
convolved model distribution of groups is shown by the dashed line in
Fig. 2; the convolved rich cluster distribution is shown by
the solid line, for comparison. The convolution flattens the model
distributions, as expected, and produces a high-velocity tail.
The rms peculiar velocities of model groups versus observations
(total sample and T-F only) are summarized in Table~1. The models are
convolved with the observational uncertainties of each sample
(total and TF) separately.

A comparison between the data and the convolved model suggests that
the observed and model velocity distributions are consistent
with each other at $\sim  3\sigma$ level. A K-S test of the velocity
distribution indicates that the model is consistent with the data
at a significance level of $\sim 3$\% (or $\sim 70$\% if only the
T-F velocities are used). The $\Omega=1$ model
is consistent with the data at a $5$\% level
(or $\sim 46$\% for the T-F data).

The observations exhibit a long tail of high velocity groups and
clusters, to $v_{1D} \sim 3000$ km s$^{-1}$. This high velocity
tail is not consistent with the convolved models.
Since the current observational uncertainties are large, and the
effect of model convolution is strong (in fact it yields most of
the high-velocity tail), more accurate velocity data are needed in
order to further constrain the cosmological models. One effect of
large observational errors is to produce an artificial high-velocity
tail. It seems likely that more accurate cluster velocities will
yield {\it smaller values than currently suggested by the observed
high-velocity tail} of Figure 2. If, however, the high velocities
($v_{1D} \geq 2000$ km s$^{-1}$) are confirmed, with $P \geq 0.01$,
it will suggest that the above models are unlikely.
\bsn
\centerline {4. SUMMARY AND CONCLUSIONS}
\msn
We find that rich clusters of galaxies exhibit a robust,
Maxwellian distribution of peculiar velocities in the
cosmological models
studied. The distribution peaks at $v \sim 400$ km s$^{-1}$ for
the $\Omega=0.3$ model, and extends to high velocities of
$v \sim 1200$ km s$^{-1}$.
Approximately 10\% of all clusters move with high peculiar
velocities of $v \ge 700$ km s$^{-1}$. The velocity distribution
of model clusters is insensitive to the cluster selection
threshold (i.e., richness). The observed distribution of peculiar
velocities of clusters of galaxies exhibits a larger velocity tail
than seen in the model; however, due to the large observational
uncertainties, the data are consistent at a $\sim 3 \sigma$ level
with the COBE-normalized low-density CDM model predictions
when convolved with the observational uncertainties. This
model fits well other large-scale structure observations. The
observed velocity distribution is consistent with
Gaussian initial density fluctuations.

There are large uncertainties in the existing measurements of
peculiar velocities; these uncertainties do not currently allow
significant constraints of the cosmological models. More
accurate observations for groups and for clusters of galaxies are
likely to yield {\it a lower velocity tail} than suggested
by the existing observations. These observations of peculiar
velocities can help to constrain cosmological models and to
test whether the initial density field was Gaussian.

\bsn
\bsn
It is a pleasure to acknowledge NCSA for
use of their Convex-3880 supercomputer.
We thank J. Goodman, J.R. Gott, J.P. Ostriker,
D.N. Spergel and R. Wijers for useful discussions.
This work was supported by
NASA grant NAGW-2448 and NSF grants
AST90-20506, AST91-08103 and AST93-15368.

\vfill\eject
\centerline {REFERENCES}
\msn
Aaronson, M., Bothun, G., Mould, J., Huchra, J., Schommer, R.A., \&
Cornell, M.E. 1986, ApJ, 302, 536
\smallskip
\refset
Bahcall, N.A. \& Soneira, R.M. 1983, ApJ, 270, 20
\smallskip
\refset
Bahcall, N.A. \& Soneira, R.M. 1984, ApJ, 277, 27
\smallskip
\refset
Bahcall, N.A., Soneira, R.M., \& Burgett, W. 1986, ApJ, 311, 15
\smallskip
\refset
Bahcall, N.A., \& Cen, R. 1992, ApJ, 398, L81
\smallskip
\refset
Bahcall, N.A., \& Cen, R. 1993, ApJ, 407, L49
\smallskip
\refset
Bertschinger, E., \& Dekel, A. 1989, ApJ, 336, L5
\smallskip
\refset
Burstein, D., Davies, R.L., Dressler, A., Faber, S.M., Lynden-Bell, D.,
Terlevich, R., \& Wegner, G. 1987, in "Galaxy Distances and Deviations
from Universal Expansion," ed. B. Madore \& R.B. Tulley
(Dordrech: Reidel), 123
\smallskip
\refset
Cen, R.Y. 1992,  ApJS, 78, 341
\smallskip
\refset
Cen, R.Y., \& Ostriker, J.P. 1992, ApJ, 399, L113
\smallskip
\refset
Cen, R.Y., Gnedin, N., \& Ostriker, J.P. 1993, ApJ, 417, 387
\smallskip
\refset
Dekel, A. 1994, ARAA, 32
\smallskip
\refset
Dressler, A., Faber, S.M., Burstein, D., Davies, R., Lynden-Bell, D.,
Terlevich, R., \& Wegner, G. 1987, ApJ, 313, L37
\smallskip
\refset
Faber, S.M., Wegner, G., Burstein, D., Davies, R.L., Dressler, A.,
Lynden-Bell, D., \&  Terlevich, R.J. 1989, ApJS, 69, 763
\smallskip
\refset
Gramann, M., 1988, MNRAS, 234, 569
\smallskip
\refset
Gramann, M., Cen, R.Y., \& Bahcall, N.A, 1994, ApJ (to be submitted)
\smallskip
\refset
Kofman, L., Gnedin, N., \& Bahcall, N.A, 1993, ApJ, 413, 1
\smallskip
\refset
Klypin, A.A., \& Kopylov, A.I. 1983, Soviet Astron. Lett., 9, 41
\smallskip
\refset
Lauer, T.R., \&  Postman, M. 1994, ApJ, 425, 418
\smallskip
\refset
Maddox, S.J., Efstathiou, G., Sutherland, W., \& Loveday, J. 1990,
MNRAS, 242, 43
\smallskip
\refset
Mathewson, D.S,, Ford, V.L., \& Buchhorn, M. 1992, ApJS, 81, 413
\smallskip
\refset
Mould, J.R., Staveley-Smith, L., Schommer, R.A., Bothun, G.D.,
Hall, P.J., Han, M., Huchra, J.P., Roth, J.,
Walsh, W., \& Wright, A.E. 1991, ApJ, 383, 467
\smallskip
\refset
Mould, J.R., Akeson, R.L., Bothun, G.D., Han, M., Huchra, J.P.,
Roth, J., \& Schommer, R.A. 1993, ApJ, 409, 14
\smallskip
\refset
Ostriker, J.P. 1993, ARAA, 31, 689
\smallskip
\refset
Peacock, J.A., \& West, M. 1992, MNRAS, 259, 494
\smallskip
\refset
Postman, M., Huchra, J., \& Geller, M. 1992, ApJ, 384, 404
\smallskip
\refset
Ramella, M., Geller, M., \& Huchra, J. 1989, ApJ, 344, 57
\smallskip
\refset
Rubin, V.C., Thonnard, N., Ford, W.K., \& Roberts, M.S. 1976, AJ, 81, 719
\smallskip
\refset
Smoot, G.F., Gorenstein, M.V., \& Muller, R.A., 1977, Phys. Rev. Lett.,
39, 898
\smallskip
\refset
Smoot, G.F., \etal 1992, ApJ, 396, L1
\vfill\eject
\noi
\centerline {FIGURE CAPTIONS}
\bsn

FIGURE 1. The 3-D peculiar velocity distribution of
clusters (solid line) and groups of galaxies (dashed line)
in the low-density CDM model. The Maxwellian distribution
for the clusters is represented by the dotted line.
(a) Differential velocity distributions;
(b) Integrated distributions.

\bsn

FIGURE 2. Comparison of observations and model predictions.
The observed velocity distributions (in 1-D velocities)
of groups of galaxies as determined from Tully-Fisher (TF) and
$D_n - \sigma$ distance-indicators (stars and open circles,
respectively), and for the combined sample (dark circles with
$\sqrt N$ uncertainties indicated) are presented.
The model 1-D group velocity distribution (dotted line), and its
convolved distribution (dashed line; convolved with the observational
velocity uncertainties) are shown. The convolved distribution
of simulated rich clusters (solid line) is also shown,
for comparison. The observations should be compared with the
convolved model group simulations (dashed line).
(a) Differential velocity distributions;
(b) Integrated distributions.

\par\vfill\eject

\bye